\documentclass[reprint,superscriptaddress,amsmath,amssymb,aps,prl]{revtex4-1}

\usepackage{graphicx}

\begin{document}

\title{Phonon lasing from optical frequency comb illumination of a trapped ion}

\author{Michael Ip}
\email{michaelip9228@physics.ucla.edu}
\affiliation{University of California Los Angeles}
\author{Anthony Ransford}
\affiliation{University of California Los Angeles}
\author{Andrew M. Jayich}
\affiliation{University of California Santa Barbara}
\author{Xueping Long}
\affiliation{University of California Los Angeles}
\author{Conrad Roman}
\affiliation{University of California Los Angeles}
\author{Wesley C. Campbell}
\affiliation{University of California Los Angeles}

\date{\today}

\begin{abstract}
An atomic transition can be addressed by a single tooth of an optical frequency comb if the excited state lifetime ($\tau$) is significantly longer than the pulse repetition period ($T_\mathrm{r}$). In the crossover regime between fully-resolved and unresolved comb teeth ($\tau \lessapprox T_\mathrm{r}$), we observe Doppler cooling of a pre-cooled trapped atomic ion by a single tooth of a frequency-doubled optical frequency comb. We find that for initially hot ions, a multi-tooth effect gives rise to lasing of the ion's harmonic motion in the trap, verified by acoustic injection locking.  The gain saturation of this phonon laser action leads to a comb of steady-state oscillation amplitudes, allowing hot ions to be loaded directly into the trap and laser cooled to crystallization despite the presence of hundreds of blue-detuned teeth.
\end{abstract}

\maketitle

The outermost valence electrons in atomic ions are tightly bound to the electron-shielded nucleus ($q_\mathrm{eff} \gtrapprox +2e$), and therefore typically require laser light in the ultraviolet (UV) for Doppler cooling \cite{Wineland1975proposed,Hansch1975cooling,Neuhauser1978optical,Wineland1978radiation}.  While continuous-wave (cw) laser light is challenging to produce in the UV, mode-locked (ML) lasers can be more easily converted to short wavelengths due to their high instantaneous power.  This makes ML lasers attractive alternatives for working with ions, but they have more complicated spectra (optical frequency combs) that can lead to deleterious effects such as the amplification of motion instead of damping.  To maximize the scattering force from the ML laser, the excited state lifetime of the cycling transition ($\tau$) should be comparable to the repetition period of the laser ($T_\mathrm{r}$) \cite{Blinov2006broadband}.  While this makes standard, tunable ML lasers ($T_\mathrm{r} \approx (100\mbox{ MHz})^{-1}$) well-aligned for manipulating laser-coolable trapped ion species ($\tau \approx 10 \mbox{ ns}$), it violates the condition necessary for being in the fully-resolved tooth limit ($\tau \gg T_\mathrm{r}$).  It is an open question whether the tooth visibility achievable with these lasers will be sufficiently robust for the cooling to overcome the amplification from the broad spectrum, particularly for initially hot ions.

Nonetheless, the promise of high UV power has already led to explorations of ML lasers as a means to laser cool species requiring light in the UV \cite{Blinov2006broadband,Kielpinski2006laser,DavilaRodriguez2016doppler,Jayich2016direct}.   In the ``broadband laser cooling'' regime ($\tau < T_\mathrm{r}$), Doppler cooling of a pre-cooled ion in a linear Paul trap was observed and presented with a simple theoretical model based on the shape of the single-pulse spectrum of a ML laser (as opposed to a frequency comb) \cite{Blinov2006broadband}.  The cooled radial mode of motion in that work reached a minimum temperature of $1\mbox{ K}$ and the axial mode reached a temperature 5 times lower than the radial mode. This is lower than the minimum temperature predicted by the single-pulse spectrum ($T_\mathrm{min}=3.4\mbox{ K}$), and we show here that even in this regime ($\tau \approx \frac{1}{4}T_\mathrm{r}$), comb tooth effects may have been responsible for these lower temperatures.  Pre-cooled trapped ions have also been laser cooled in the resolved-tooth regime ($\tau > T_\mathrm{r}$) using a high-repetition-rate ($1/T_\mathrm{r} = 373 \mbox{ MHz} $) frequency comb \cite{DavilaRodriguez2016doppler}.  The ML laser in that work was unable to crystallize initially hot ions, and several improvements were suggested for achieving this goal, including further shortening the repetition period of the laser to suppress scattering from other teeth.

In this Letter, we show that even when the laser repetition period is significantly longer than the atomic excited state lifetime, coherent effects permit practical cooling and crystalization of hot ions. We report trap loading and frequency comb laser cooling of trapped ions in the crossover regime between the well-resolved and unresolved comb tooth limits ($\tau \lessapprox T_\mathrm{r}$ ), where 78\% of the excited state population decays between consecutive pulses. We find that, instead of heating the ion out of the trap, the broad frequency comb gives rise to stable phonon lasing of the ion's motion \cite{Vahala2009a}, an interpretation that is confirmed by the observation of injection locking of the ion's amplified mechanical oscillation \cite{Knunz2010injection}. Irrespective of the sign of detuning for the comb tooth closest to resonance, saturable vibrational lasing leads to a comb of stable, fixed-point oscillation amplitudes maintained by cooperative effects between red and blue detuned comb teeth.  As a practical matter, the saturation of the phonon lasing mitigates the motional amplification of the blue-detuned teeth, allowing direct loading of ion traps with broad spectral coverage.

\begin{figure}
\begin{center}
\includegraphics[width=1.0\columnwidth]{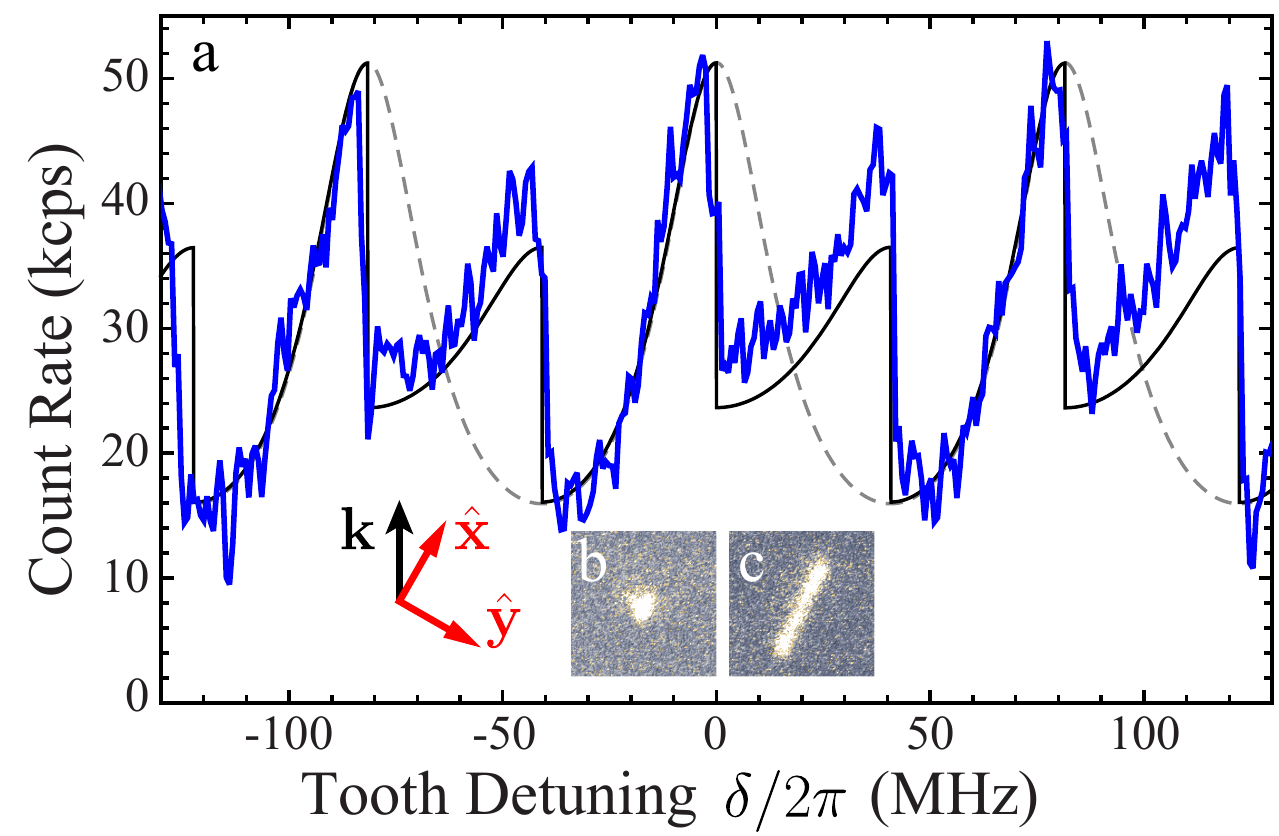}
\end{center}
\caption{Fluorescence spectrum (in kilocounts per second) from a single trapped ion illuminated by an optical frequency comb.  When the near-resonant tooth is red detuned, the ion is Doppler cooled near the ground state (b) and the spectrum (a) follows the rest-frame lineshape from Eq.~\ref{eqn:barecomb} (dashed, gray).  When the near-resonant tooth is blue detuned, the ion oscillates with a fixed amplitude (c) and the fluorescence shows clear departure from the natural rest-frame resonance shape.  The solid black curve is from Eq.~\ref{eqn:Gamma} with oscillation amplitude $x_0=0$ or $x_0=4.9\mbox{ }\mu\mbox{m}$ when the near-resonant comb tooth is red or blue detuned, respectively.}
\label{fig:WideCombScan}
\end{figure}

Single ${}^{174}\mathrm{Yb}^+$ (or ${}^{171}\mathrm{Yb}^+$) ions for this work are confined in a radio-frequency (rf) Paul trap with oblate spheroidal symmetry \cite{Yoshimura2015creation}, driven by a sinusoidal voltage at $\Omega_\mathrm{rf} = 2 \pi \times 48.535\mbox{ MHz}$.  The secular frequencies of motion are typically $(\omega_x,\omega_y,\omega_z) \approx 2 \pi \times (550,500,920)\mbox{ kHz } (\pm 10 \mbox{ kHz})$. We illuminate the ions with laser beams that are confined to the $xy$-plane and $\omega_x$ is the frequency of the phonon mode that lases most frequently due to the fact that it is closest to parallel to the laser propagation direction ($\Delta \theta = 23^\circ \pm 2^\circ$, see Fig.~\ref{fig:WideCombScan}). Fluorescence emitted in the $-z$ direction is collected by an imaging system and recorded on a photon-counting photomultiplier tube (PMT) and intensified CCD camera through a $369 \mbox{ nm}$ bandpass filter.

The ML laser for this work is a commercial picosecond Ti:sapphire laser oscillator \footnote{SpectraPhysics Tsunami} with a repetition rate of $f_\mathrm{r} = 1/T_\mathrm{r} = 81.553\mbox{ MHz}$.  The center frequency of the laser is set near $405.645\mbox{ THz}$, and the output is frequency doubled via single-pass through  a $0.8\mbox{ cm}$ LBO crystal cut for Type I phase matching for SHG of 760\mbox{ nm} light, generating an average UV power of around $9\mbox{ mW}$ at $\lambda \approx 369.5\mbox{ nm}$.  The laser bandwidth ($> 10 \mbox{ GHz}$) far exceeds the natural linewidth ($\gamma \equiv 1/\tau = 2 \pi \times 19.7\mbox{ MHz}$) and Zeeman splitting ($< 5 \mbox{ MHz}$) of the ${}^2P_{1/2} \leftrightarrow {}^2S_{1/2}$ transition in $\mathrm{Yb}^+$.  The peak ion fluorescence observed indicates that the optical power per comb tooth at the ion is $\approx 1.4 \mbox{ }\mu\mbox{W}$.  For all of the experiments reported here, we simultaneously illuminate the trapped ion with the mode-locked laser at $369.5\mbox{ nm}$ and a cw repump laser at $935\mbox{ nm}$ that has no observable direct mechanical effect on the ion.

For these parameters, the probability of excited state decay between pulses $(1 - \mathrm{e}^{-T_\mathrm{r}/\tau})$ is $78$\%, and each decay makes the ion insensitive to the optical phase of the next pulse, suggesting that the comb teeth will not be well resolved.  Quantitatively, the quasi-steady-state scattering rate for an atom at rest illuminated by a resonant comb of uniformly intense teeth ($\tau_\mathrm{pulse} \ll T_\mathrm{r}$) is given by \cite{Felinto2003coherent,Ilinova2011doppler,Aumiler2012simultaneous}
\begin{equation}
\Gamma_\mathrm{comb}(\delta) = \frac{1}{T_\mathrm{r}}\frac{\sin^2\left( \frac{\theta}{2} \right) \sinh \left( \frac{T_\mathrm{r}}{2 \tau}\right)}{\cosh \left( \frac{T_\mathrm{r}}{2 \tau}\right)- \cos^2\left( \frac{\theta}{2} \right)\cos (\delta T_\mathrm{r})},\label{eqn:barecomb}
\end{equation}
where $\delta$ is the angular detuning of a reference tooth from resonance ($\delta \equiv \omega_\mathrm{tooth} - \omega_\mathrm{atom}$) and $\theta$ is the pulse area, defined as the integral of the instantaneous Rabi frequency for a single pulse, $\theta \equiv \int\mathrm{d}t \Omega(t)$.

We quantify the relative importance of considering comb structure vs.\ the broadband (single-pulse) spectrum by the tooth visibility $V \equiv (\Gamma_\mathrm{max} - \Gamma_\mathrm{min}) / (\Gamma_\mathrm{max} + \Gamma_\mathrm{min})$, where $\Gamma_\mathrm{min}$ and $\Gamma_\mathrm{max}$ are the minimum and maximum scattering rates as the comb is scanned over a range of $\Delta f = 1/T_\mathrm{r}$.  From Eq.~\ref{eqn:barecomb}, we find
\begin{equation}
V = \cos^2\left( \frac{\theta}{2}\right) \mathrm{sech}\left( \frac{T_\mathrm{r}}{2 \tau}\right),\label{eqn:Visibility}
\end{equation}
which shows that phase-coherent, inter-pulse effects persist remarkably far into the $\tau < T_\mathrm{r}$ regime, tracking the decay of the off-diagonal elements of the density matrix (i.e. the coherences) more closely than the populations.  For instance, the lifetime-limited ($\theta \!\!\rightarrow \!\!0$) tooth visibility in this work is $V=0.77$ despite the fact that more than $3/4$ of the excited state population decays between pulses.

Figure \ref{fig:WideCombScan}(a) shows the $369.5\mbox{ nm}$ fluorescence collected from a trapped ion as the teeth of the optical frequency comb illuminating it are scanned. The dashed gray curve is given by Eq.~\ref{eqn:barecomb}, and has been adjusted for the overall vertical scale (with no offset) and is shown for $\theta = 0.38\, \pi$, which has a tooth visibility of $V=0.53$. The data are taken for $100 \mbox{ ms}$ of illumination per point so that the ion's motional temperature reaches steady-state for each frequency shown, and background counts $(\approx 170 \mbox{ kcps})$ have been subtracted by taking the difference between the PMT signal with and without the $935 \mbox{ nm}$ repump laser.

\begin{figure}
\begin{center}
\includegraphics[width=1.0\columnwidth]{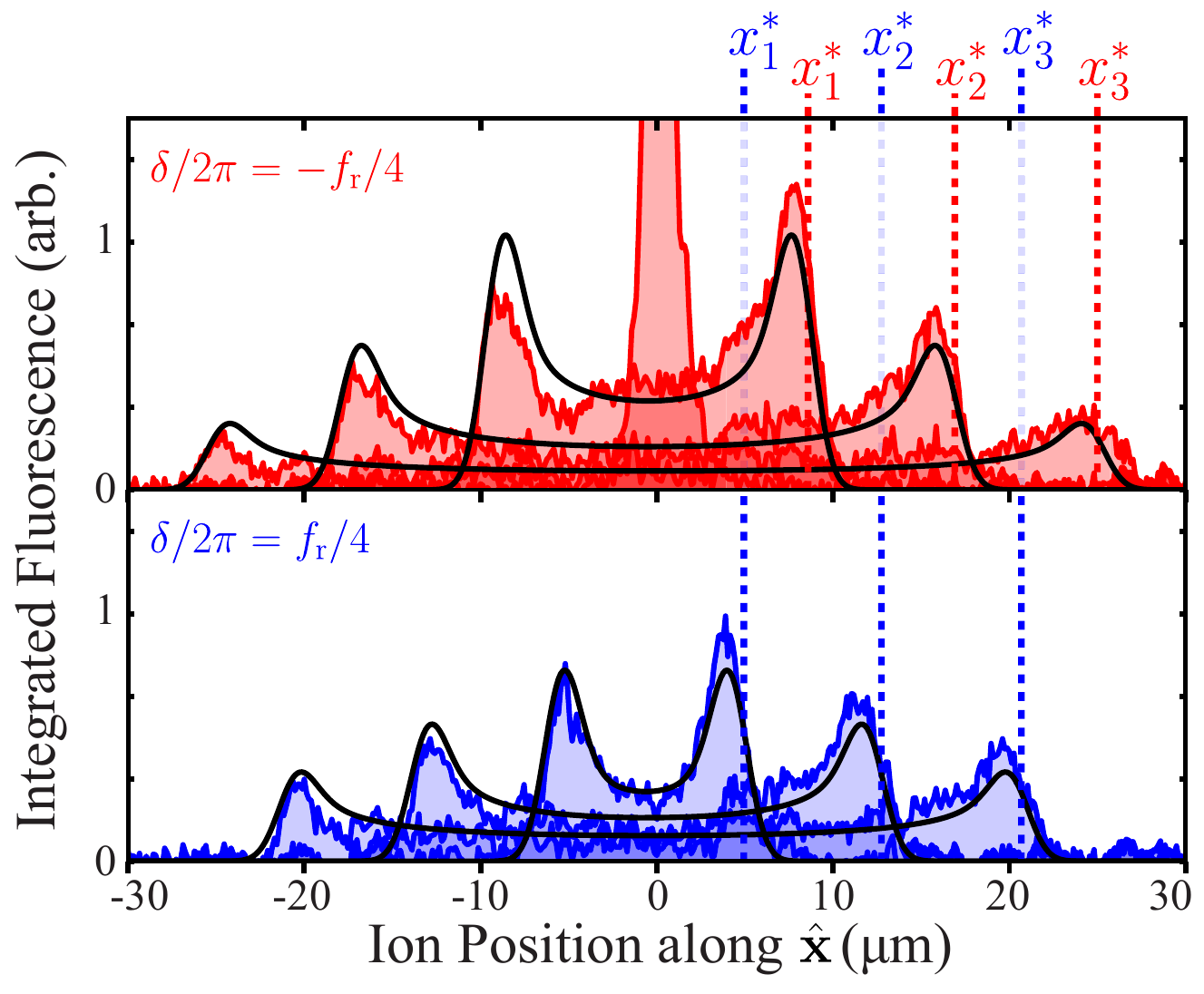}
\end{center}
\caption{Integrated spatial images of fluorescence from an ion illuminated by an optical frequency comb whose nearest-resonant tooth is detuned to the red (upper) or blue (lower) of rest-frame resonance.  Multiple fixed points are shown (as individual traces) for each case.  Hysteresis associated with the ion's initial energy determines which fixed point the ion finds.  The dashed lines indicate the fixed-point oscillation amplitudes extracted from fitting to classical harmonic oscillator distributions (solid black curves).}
\label{fig:ColumnSums}
\end{figure}

The fluorescence in Fig.~\ref{fig:WideCombScan}(a) tracks the ion's (power-broadened) rest-frame lineshape well when the near-resonant tooth is on the red side of atomic resonance ($\delta < 0$), which is the regime in which cw-like Doppler cooling actively cools the ion's motion. As the near-resonant tooth crosses resonance and becomes blue-detuned ($\delta>0$), the fluorescence no longer follows the rest-frame prediction of Eq.~\ref{eqn:barecomb}, and an additional peak appears. As shown in Fig.~\ref{fig:WideCombScan}(c), the ion image in this case spreads out along a principal axis of the trap with an amplitude of $x_0=(4.9 \pm 0.5) \mbox{ }\mu\mbox{m}$. This fluorescence image matches well to the classical probability distribution of a harmonic oscillator convolved with the point-spread function of our imaging system (see, \textit{e.g.} Fig.~\ref{fig:ColumnSums}), which shows that the ion is oscillating with fixed amplitude.

To model the fluorescence spectrum of an ion oscillating slowly ($\omega_x \ll \gamma$) with secular amplitude $x_0$, we calculate the average of the atomic scattering rate over an oscillation period in the trap.  Since $\Omega_\mathrm{rf}>\gamma$, we include the effects of rf micromotion as an incoherent sum over the scattering rates from secular-phase-dependent micromotion sidebands \cite{Blumel1989chaos}, with each sideband yielding fluorescence with the functional form of Eq.~\ref{eqn:barecomb}.  Defining $\xi \equiv \omega_x t$, the secular-cycle-averaged scattering rate is given by
\begin{eqnarray}
\Gamma(\delta, x_0) = \frac{1}{2 \pi} \int_0^{2 \pi} \mathrm{d}\xi \sum_n \mathrm{J}_n^2\!\left[ k x_0 \frac{q}{2} \cos(\xi)\right]  \nonumber\\
\times \Gamma_\mathrm{comb} \left(\delta + n \Omega_\mathrm{rf} + k \omega_x x_0 \sin(\xi) \right)\label{eqn:Gamma}
\end{eqnarray}
where $q\approx 2 \sqrt{2}\omega_x/\Omega_\mathrm{rf}$ is the Mathieu $q$-parameter for this normal mode, $k$ is the projection of the laser's k-vector on the oscillation axis, and $\mathrm{J}_n[\alpha]$ is the $n$th Bessel function of the first kind. The black curve in Fig.~\ref{fig:WideCombScan} shows the scattering rate predicted by this model (scaled for the overall signal height) with a constant oscillation amplitude of $x_0=4.9 \mbox{ }\mu\mbox{m}$ when the nearest-resonant tooth is on the blue side of resonance and $x_0=0$ on the red side.

A similar type of stable, laser-driven, large amplitude motion has been reported for an ion illuminated by two colors of cw laser light with one beam red-detuned and the other, weaker beam blue-detuned, and was shown to be a phonon laser \cite{Vahala2009a}.  In the case studied here, the ML comb teeth all have nearly the same strength and there are hundreds, as opposed to two frequencies. We present a model suited to these features below that describes the phonon laser behavior that arises.

To model the ion-ML laser interaction, we work with cycle-averaged expressions for the secular energy $E\equiv \frac{1}{2}m \omega_x^2 x_0^2$, amplitude damping coefficient $\beta(E)$, and stochastic heating rate $S(E)$ from spontaneous emission and the randomness in absorption.  The rate of energy transfer from the optical comb to the ion's motion (the net mechanical amplification power) is given by
\begin{equation}
\frac{\mathrm{d}E}{\mathrm{d}t} = -\beta(E) \frac{2E}{m} + S(E). \label{eqn:PowerBalance}
\end{equation}
The stochastic heating power can be calculated from the total scattering rate, Eq.~\ref{eqn:Gamma}: $S(E) = (1 + \zeta)\frac{\hbar^2 |\mathbf{k}|^2}{2 m} \Gamma(\delta, x_0)$, where $\zeta = \frac{2}{5}$ is the geometric factor for dipole emission \cite{Leibfried2003quantum} and $x_0 = \sqrt{2E/m \omega_x^2}$.  The coherent (i.e.~non-stochastic) amplification power is given by $-\beta 2 E/m = \left\langle\mathbf{F}\cdot \mbox{\boldmath$v$}_\mathrm{sec}(\xi)\right \rangle_\xi$, which yields the following expression for the amplitude damping coefficient,
\begin{eqnarray}
\beta(E) = \frac{\hbar k}{2 \pi \omega x_0} \int_0^{2 \pi} \mathrm{d}\xi \sin(\xi) \sum_n \mathrm{J}_n^2\!\left[ k x_0 \frac{q}{2} \cos(\xi)\right]  \nonumber\\
\times \Gamma_\mathrm{comb} \left(\delta + n \Omega_\mathrm{rf} + k \omega x_0 \sin(\xi) \right).\label{eqn:Beta}
\end{eqnarray}

\begin{figure}
\begin{center}
\includegraphics[width=1.0\columnwidth]{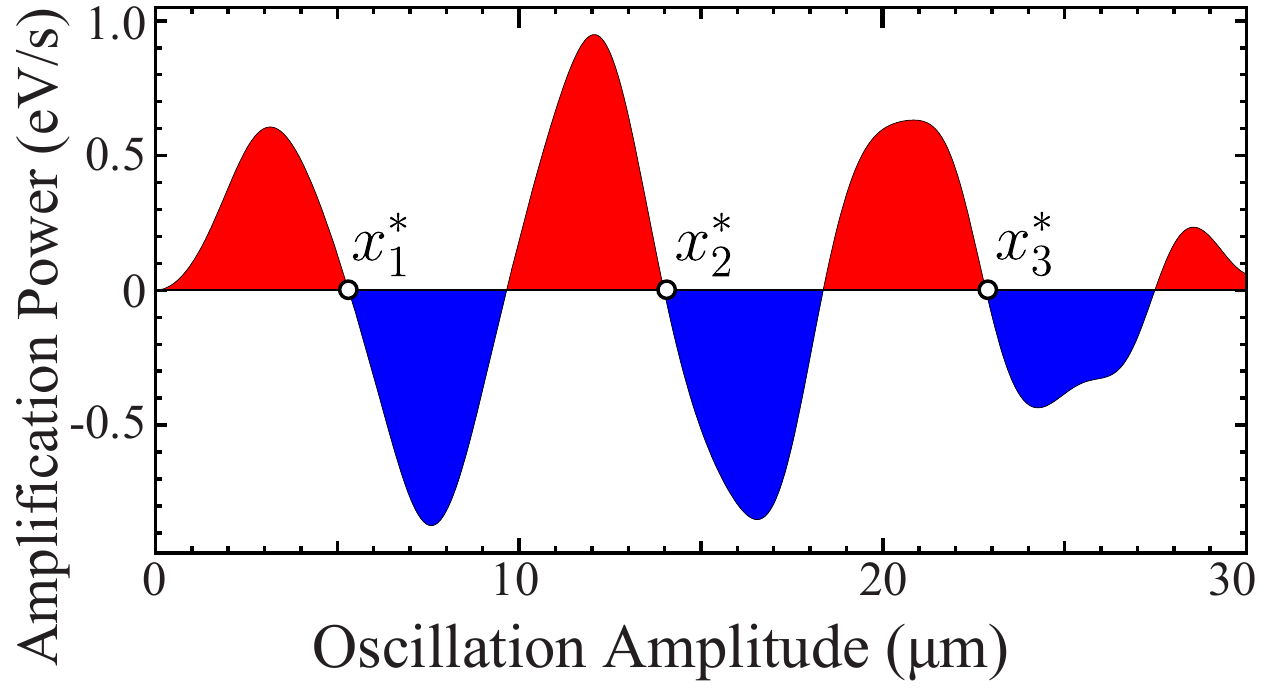}
\end{center}
\caption{Calculated net power delivered to the motion of a trapped $\mathrm{Yb}^+$ atom illuminated by an optical frequency comb as a function of the amplitude of the oscillations in the trap. The right side of Eq.~\ref{eqn:PowerBalance} is plotted for the case where the near-resonant tooth is blue-detuned with $\delta = \pi / 2 T_\mathrm{r}$.  Red (blue) shading indicates amplification (damping) of the ion's motion.  All zero-crossings are fixed points, with only those corresponding to a negative slope in this figure being stable, shown by circles.}
\label{fig:TheoryPowervsAmplitude}
\end{figure}

In the limit of small oscillations ($E \rightarrow 0$), the damping coefficient $\beta$ changes sign with $\delta$ across resonance.  If the comb tooth closest to resonance is red-detuned, $\beta>0$ and it can Doppler cool the ion's motion much like a single, cw laser, as shown in Fig.~\ref{fig:WideCombScan}(b). Eq.~\ref{eqn:PowerBalance} predicts a steady-state oscillation amplitude of $x_0^\ast = 90\mbox{ nm}$ for $\delta = -\pi/2T_\mathrm{r}$, much smaller than the resolution ($\approx 1 \mbox{ }\mu\mbox{m}$) of our imaging system.  This oscillation energy corresponds to $E/k_\mathrm{B} = 980 \mbox{ }\mu\mbox{K}$, less than a factor of two higher than the cw Doppler limit predicted for the same saturation parameter ($T_\mathrm{D} = 530 \mbox{ }\mu\mbox{K}$). Thus, even with substantial spontaneous decay between pulses $(\tau \lessapprox T_\mathrm{r})$, the comb structure is sufficiently robust that the cooling effect of the nearest tooth is able to overcome the amplification and heating induced by other parts of the spectrum.

When the near-resonant comb tooth is on the blue side of the atomic resonance ($\delta > 0$), for small oscillation amplitude there is net gain for the ion's motion from the laser field ($\beta<0$).  This amplification has recently been observed spectroscopically as a line pulling mechanism \cite{Ozawa2017single}.  Figure \ref{fig:TheoryPowervsAmplitude} shows $\mathrm{d}E/\mathrm{d}t$ calculated from Eq.~\ref{eqn:PowerBalance} for this case. The frequency comb will add energy to the motion of an initially cold ion until the net power transfer vanishes, leading to a steady-state oscillation that can occur with significant amplitude, as shown in Fig.~\ref{fig:WideCombScan}(c).

In fact, multiple roots of the right side of Eq.~\ref{eqn:PowerBalance} exist for both signs of $\delta$. Those roots with a negative derivative with respect to $E$ are stable fixed points of this nonlinear differential equation, denoted here with a superscript $\ast$ (see Fig.~\ref{fig:TheoryPowervsAmplitude}) \cite{Strogatz}. Experimentally, we find that stable fixed points persist while the comb teeth are stable in fequency, but if the sign of $\delta$ is changed, a new fixed point is reached that is approximately equally likely to be lower or higher in amplitude.  By scanning $\delta$, we can regularly load and crystallize ions (including multi-ion crystals) with only the ML laser.

Figure \ref{fig:ColumnSums} shows the experimental signature of multiple fixed point solutions for the oscillation amplitude when the near-resonant tooth is red (upper) or blue (lower) detuned.  These integrated fluorescence images resemble the ``two-lobe'' shape of an image of a classical harmonic oscillator (or coherent state) probability distribution, shown as black curves.  The fits (which include convolution with our imaging system point-spread function) are used to extract the oscillation amplitude fixed points with a resolution-limited systematic uncertainty of $\pm 0.5 \mbox{ }\mu\mbox{m}$.  When the near-resonant tooth is red detuned, we find stable oscillation amplitudes $(x^\ast_1, x^\ast_2, x^\ast_3) = (8.6, 16.9, 25.0) \mbox{ }\mu\mbox{m}$, compared to the theoretical prediction of $(9.7, 18.4, 27.5)\mbox{ }\mu\mbox{m}$ from the roots of Eq.~\ref{eqn:PowerBalance}.  For the blue-detuned case, we find $(x^\ast_1, x^\ast_2, x^\ast_3) = (4.9, 12.7, 20.7) \mbox{ }\mu\mbox{m}$, with the corresponding predicted values $(5.3, 14.0, 22.9) \mbox{ }\mu\mbox{m}$ (see Fig.~\ref{fig:TheoryPowervsAmplitude}).  The measured fixed points agree with the predicted values to about $10$\%.

\begin{figure}
\begin{center}
\includegraphics[width=1.0\columnwidth]{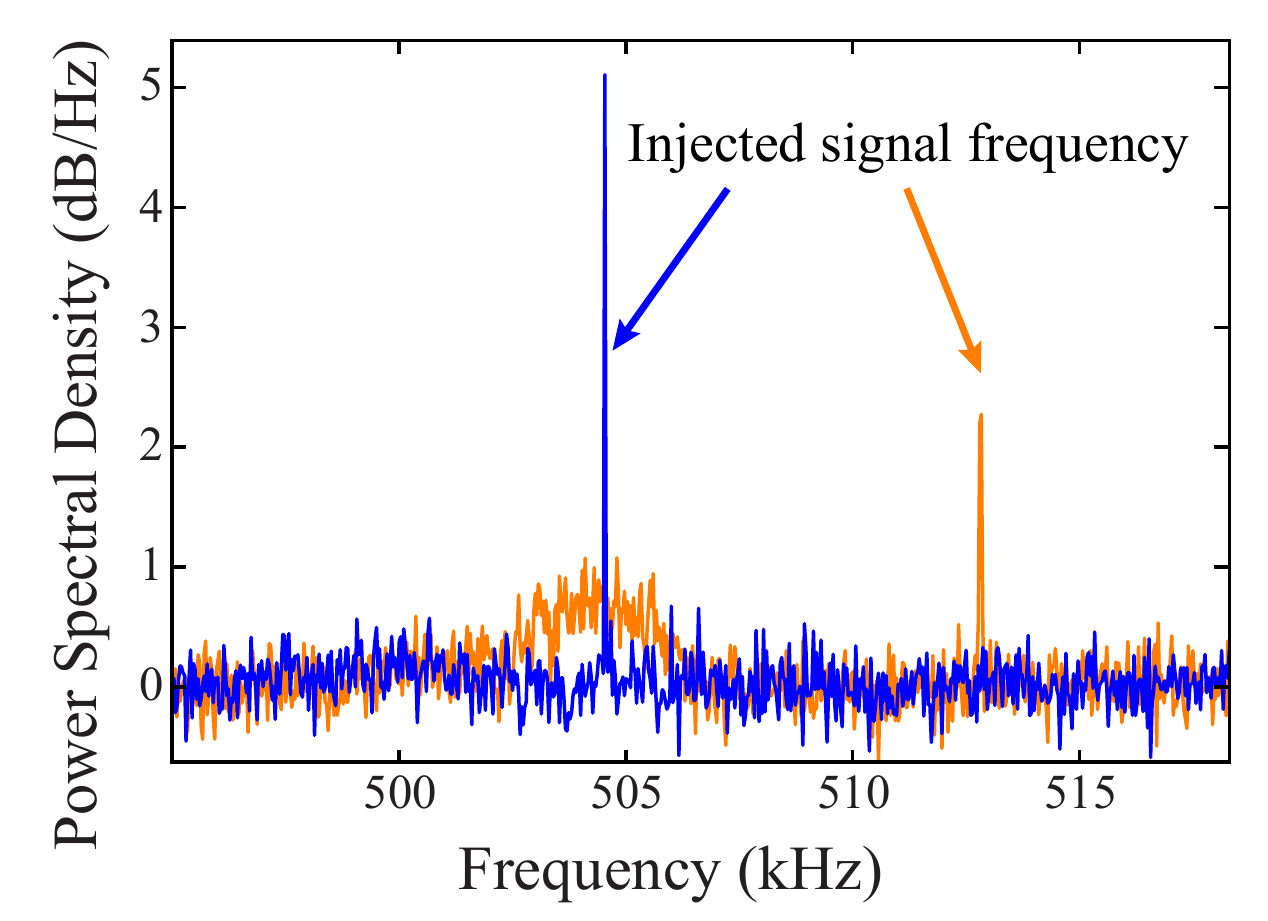}
\end{center}
\caption{Acoustic injection locking of the $y^\ast_1$ fixed point phonon laser when the near-resonant tooth is blue detuned.  The spatially-filtered frequency spectrum of the ion fluorescence is shown in the presence of a sinusoidal voltage that has been applied to one of the trap electrodes.  When the frequency of this voltage is moved from outside (orange) to within (blue) the phonon laser's gain bandwidth, it is amplified at the expense of other frequencies.}
\label{fig:InjectionLock}
\end{figure}

We further verified that the system behaves as a phonon laser amplifier by observing acoustic injection locking \cite{Knunz2010injection} for each of the first three fixed points (other than the $x_0^\ast \approx 0$ fixed point for $\delta<0$) for both signs of $\delta$.  Using an aperture in the imaging system, ion fluorescence is collected from only one of the classical turning points in space, and the photons are time-tagged \cite{Pruttivarasin2015compact}.  Figure \ref{fig:InjectionLock} shows a numerical discrete Fourier transform of the recorded photon signal in the region near the secular frequency of the ion ($\omega_y$ in this case).  A narrow peak appears at the frequency of a small sinusoidal voltage that was applied to one of the trap electrodes (the injected signal), and the broad peak is the free-running phonon laser spectrum.  When the injected oscillation frequency is within this gain bandwidth of the phonon laser (shown as a broad bump in the orange trace), almost all of the gain is dedicated to amplifying the injected signal, and oscillator locks on to the injected signal in frequency and phase.

Both the theoretical model and the observed behavior we report here show that phenomena arising from inter-pulse coherence can still dominate ML-illuminated ion dynamics in the regime where $\tau < T_\mathrm{r}$.  For the system described in Ref.~\cite{Blinov2006broadband}, the probability of excited state decay between consecutive pulses was greater than 98\%, but Eq.~\ref{eqn:Visibility} shows that the limit on tooth visibility imposed by this decay is nonetheless $V=0.27$.  In this case, our model predicts that stable fixed points will exist, with the lowest fixed point energies lying between $2 \mbox{ mK}$ and $2 \mbox{ K}$, depending upon $\delta$.  This suggests that it is possible comb tooth effects were responsible for the anomalously low temperatures that were observed in that work \cite{Blinov2006broadband}.

Since blue-detuned comb teeth do not pose an intractable problem for working with trapped ions, optical frequency combs may prove to be useful and robust tools for a variety of currently difficult species. In the case of $\mathrm{He}^+$, which is an attractive spectroscopic subject \cite{Herrmann2009feasibility}, $1\mbox{ mW}$ frequency comb light sources may soon become available at $\lambda \approx 61\mbox{ nm}$ \cite{Porat2017phase}, and 2-photon transitions could be used for slow Doppler cooling in the limit of low intensity \cite{Jayich2016direct}. Direct comb cooling of helium ions may then enable work with ${}^3\mathrm{He}^+$, the lightest atomic cation with an electron (and therefore the most difficult to sympathetically cool with another ion), where the ground state hyperfine splitting of $8.67\mbox{ GHz}$ \cite{Schuessler1969hyperfine} would easily be spanned by a comb for repumping and spectroscopy. We have tested one possible hyperfine repumping scheme for $I=1/2$ experimentally by laser cooling and repumping ${}^{171}\mbox{Yb}^+$ with only the ML laser, similar to the case reported for ${}^{25}\mathrm{Mg}^+$ \cite{DavilaRodriguez2016doppler}.

\begin{acknowledgments}
The authors acknowledge discussions with Boris Blinov and thank Eric Hudson for comments on the manuscript.  This work was supported by the U.S. Army Research Office under Grant No. W911NF-15-1-0261.  Initial work by A.M.J. and X.L. was supported by the NSF CAREER Program under award No. 1455357.
\end{acknowledgments}

M.I. and A.R. contributed equally to this work. The manuscript first draft was written by M.I. and W.C.C.
\bibliography{CombCoolingPhononLasing}
\end{document}